\documentclass[twocolumn]{elsart}

\usepackage{natbib}

\usepackage{graphics}
\usepackage{graphicx}

\usepackage{amssymb}

\begin{document}

\begin{frontmatter}



\title{The XENON Dark Matter Search Experiment}

\author[columbia]{E. Aprile\corauthref{cor}}, \corauth[cor]{Presented at the 6th
  UCLA Symposium on \textit{Sources and Detection of Dark Matter and Dark Energy in the Universe.}} \ead{age@astro.columbia.edu}
\author[columbia]{K.L. Giboni}, \author[columbia]{P. Majewski}, 
\author[columbia]{K. Ni}, \author[columbia]{M. Yamashita}, 
\author[brown]{R. Gaitskell}, \author[brown]{P. Sorensen},
\author[brown]{L. DeViveiros}, \author[florida]{L. Baudis},
\author[lawrence]{A. Bernstein},
\author[lawrence]{C. Hagmann}, \author[lawrence]{C. Winant}, \author[princeton]{T. Shutt},
\author[princeton]{J. Kwong}, \author[rice]{U. Oberlack},
\author[yale]{D. McKinsey}, \author[yale]{R. Hasty}

\address[columbia]{Physics Department and Columbia Astrophysics Laboratory, Columbia University, New
  York, NY 10027}
\address[brown]{Physics Department, Brown University, Providence, RI 02912}
\address[florida]{Physics Department, University of Florida, Gainesville, FL 32611}
\address[lawrence]{Lawrence Livermore National Laboratory, 7000 East Ave.,
  Livermore, CA 94550}
\address[princeton]{Physics Department, Princeton University, Princeton, NJ 08544}
\address[rice]{Physics Department, Rice University, Houston, Texas 77251}
\address[yale]{Physics Department, Yale University, New Haven, CT 06520}

\begin{abstract}
The XENON experiment aims at the direct detection of dark
matter in the form of WIMPs (Weakly Interacting Massive Particles) via their
elastic scattering off Xe nuclei. A fiducial mass of 1000 kg, distributed in ten independent liquid xenon time projection chambers(LXeTPCs) will be used to probe the
lowest interaction cross section predicted by SUSY models. The TPCs are operated
in dual (liquid/gas) phase, to allow a measurement of nuclear recoils down to 16
keV energy, via simultaneous detection of the ionization, through secondary
scintillation in the gas, and primary scintillation in the liquid. The
distinct ratio of primary to secondary scintillation for nuclear
recoils from WIMPs (or neutrons), and for electron recoils from background, is
key to the event-by-event discrimination capability of XENON. A dual phase xenon
prototype has been realized and is currently being tested, along with other
prototypes dedicated to other measurements relevant to the XENON program. As
part of the  R\&D phase, we will realize and move underground a first XENON
module (XENON10) with at least 10 kg fiducial mass to measure the background
rejection capability and to optimize the conditions for continuous and stable
detector operation underground.  We present some of the results from the ongoing
R\&D and summarize the expected performance of the 10 kg experiment, from Monte
Carlo simulations. The main design features of the 100 kg detector
unit(XENON100), with which we envisage to make up the 1 tonne sensitive mass of
XENON1T will also be presented.
 
\end{abstract}


\end{frontmatter}

\section{Introduction}
\label{intro}

Combined analyses of the latest observational data continue to provide
compelling evidence for a significant cold dark matter component in the
composition of the Universe \citep{Freedman:03}. While the composition of dark
matter is still unknown, WIMPs are a particularly well-motivated
dark matter candidate.  \citep{Munoz:03}. Numerous direct and indirect detection experiments
have been on-going for decades, and a number of newly proposed
projects are being developed \citep{Chardin:03}. 

The contradicting results between an annual modulation signal from the DAMA
experiment \citep{DAMA:03} and the lowest exclusion limits set by the CDMS
experiment \citep{CDMS:04} requires a new generation of dark matter direct
search experiments with the sensitivity pushed several orders of magnitude higher.

Efficient and redundant background rejection schemes are a key
requirement for these future WIMP experiments, along with the capability
to sense nuclear recoil energy depositions as low as a few keV. One of
the emerging technologies is to use liquid xenon (LXe) as the sensitive
target.  Particle interactions produce both scintillation and
ionization signals in LXe, and 
the distinct ratio of these two signals for nuclear and electron recoil events
provides a powerful and efficient background rejection\citep{Yamashita:03}.  

\begin{figure}[htbp] 
  \begin{center} 
   \includegraphics[width=7cm]{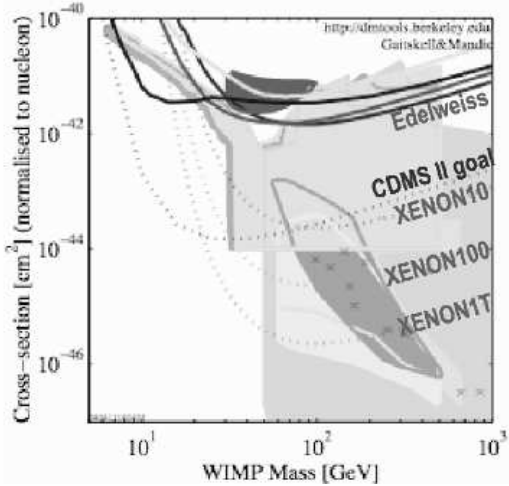} 
   \caption{Current direct detection dark matter limits, along with the projected sensitivity
   of XENON10, XENON100 and XENON1T.}   
   \label{sensitivity} 
  \end{center} 
\end{figure}

To achieve an increase in sensitivity to $\sim$
10$^{-46}$ cm$^2$ requires a fiducial target mass on the order of 1 tonne, with less than about 10 background events per year. The
sensitivity of the current cryogenic experiments will ultimately be limited by the
mass of available target, given the technical challenges of scaling well
beyond $\sim$ 50 kg. The availability of LXe in large quantities
at a reasonable cost, opens the possibility for experiments on the tonne scale. For
the XENON \citep{XENON:01} experiment, we have proposed a mass of 1 tonne
distributed in ten LXeTPCs. The design goals are 99.5\% background rejection
efficiency, achieved by simultaneous measurement of
ionization and scintillation signals and an energy threshold of about 16 keV (at 
which the detector is fully efficient for dark matter detection). Event
localization in 3-D and the use of a LXe self-shield provide additional
discrimination power. Fig. \ref{sensitivity} shows current direct
detection dark matter limits along with the projected
sensitivity for XENON in its various implementation phases.

\section{Baseline Detector Design}
\begin{figure}[htbp] 
  \begin{center} 
   \includegraphics[width=5cm]{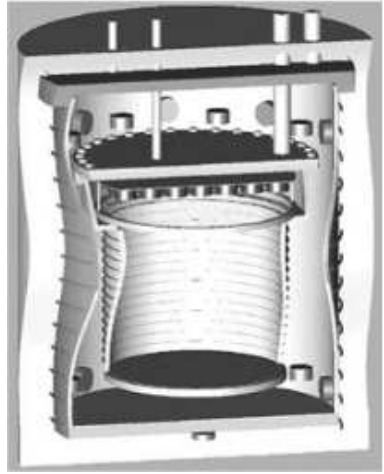} 
   \caption{Schematic view of the XENON 100 dual phase detector.}   
   \label{xe100} 
  \end{center} 
\end{figure}

Fig.\ref{xe100} schematically shows the design of the detector proposed as unit
module for the XENON experiment. It is a dual phase TPC, with the active LXe
volume defined by a 50 cm diameter CsI photocathode immersed in the
liquid, at about 30 cm from the first of three wire grids defining a gas
proportional scintillation region. An array of 85 two-inch diameter, UV sensitive PMTs
located above the grids, is used to detect both primary and secondary light. The
baseline design uses Hamamatsu R9288 Photomultiplier Tubes (PMTs), as in
the prototype discussed below, but with selected materials for low radioactivity.

The TPC is enclosed in a leak-tight cylindrical structure made of
PTFE and OFHC. The PTFE is used as effective UV light reflector \citep{Yamashita:04}
and as electrical insulator. The fraction of direct light heading
downward will be efficiently detected with the CsI photocathode
\citep{Aprile:94}. The whole structure is immersed in a bath of LXe, serving
as active veto shield against background. The LXe for shielding
is contained in a double wall vacuum cryostat, made of stainless
steel and is cooled by a pulse tube refrigerator. An array of 64
PMTs are mounted on the walls of the cryostat, fully immersed in
LXe to detect the direct scintillation light from the shield. 
With both target and shield Xe volumes kept at the same
temperature and pressure, the thickness of the vessel enclosing
the TPC can be minimized. The amount of active Xe in the TPC is
about 180 kg. With  fiducial volume cuts applied for background
reduction to the level required for a sensitive WIMP search,
the active target is reduced to about 120 kg, hence we refer to this unit module
as XENON100. Clearly, if the
dominant radioactivity from the PMTs is reduced, we will recover
a larger fraction of the active target for a WIMP search.

An event in the XENON TPC will be characterized by three signals
corresponding to detection of direct scintillation light, proportional
light from ionization electrons and CsI photoelectrons. Since electron diffusion in LXe is
small, the proportional scintillation pulse is produced in a small spot
with the same X-Y coordinates as the interaction site, allowing 2D
localization with an accuracy of 1 cm. With the more precise Z
information from the drift time measurement, the 3D  event localization
provides an additional background discrimination via fiducial volume
cuts. The simulated detection efficiency of the primary scintillation light is
about 5 p.e./keV for the XENON 100 detector.

\section{Current Results from the XENON R\&D Phase}
\label{}
The XENON R\&D program is being carried out with various prototypes dedicated to test several feasibility aspects of the proposed concept, and to measure the relevant detector characteristics. Here we limit the discussion to the results obtained to-date with a dual phase xenon prototype with $\sim$ 3 kg of active mass. The primary 
scintillation light ($S1$) from the liquid, and the secondary scintillation
light ($S2$) from the ionization electrons extracted into
the gas phase, are detected  by an array of seven PMTs, operating in the cold gas above the liquid. Excellent discrimination
between alpha particles and gamma-ray events was achieved with this prototype
chamber. We are currently upgrading the detector light readout system to be able
to measure low energy nuclear recoils.
 
\subsection{Set Up and DAQ System}
\begin{figure}[htbp] 
  \begin{center} 
   \includegraphics[width=6cm]{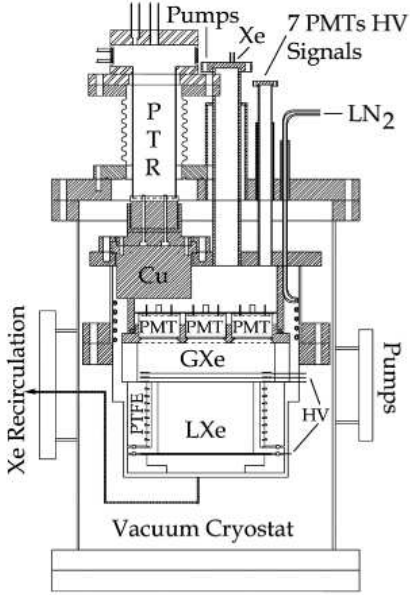} 
   \caption{Schematic drawing of the dual phase prototype.}   
   \label{prototype} 
  \end{center} 
\end{figure}

\begin{figure}[htbp] 
  \begin{center} 
   \includegraphics[width=6cm]{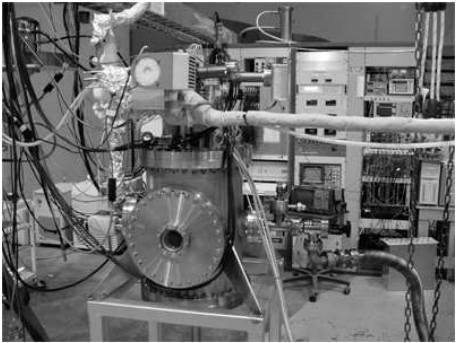} 
   \caption{The detector integrated with the vacuum cryostat, refrigerator, gas/recirculation and DAQ systems.}   
   \label{setup} 
  \end{center} 
\end{figure}

A drawing of the TPC prototype is shown in Fig. \ref{prototype} while
the photo of Fig. \ref{setup} shows the detector integrated with the vacuum
cryostat, refrigerator, gas/recirculation system. and electronics for control and data taking. The detector is
cooled by a Pulse Tube Refrigerator (PTR), optimized for LXe temperature.

The sensitive volume of the TPC (7.7$\times$7.7$\times$5.0 cm$^3$) is
defined by PTFE walls and grids with high optical transmission, made of Be-Cu wires with a pitch of 2 mm
and 120$\mu$m diameter. Negative HV is applied to the bottom grid, used as cathode. Grids on the top close the charge drift region in the liquid
and with appropriate biasing, create the amplification region for gas proportional scintillation.
Shaping rings located outside of the PTFE
walls and spaced 1.5 cm apart, are used to create a uniform electric field for
charge drift.

\begin{figure}[htbp] 
  \begin{center} 
   \includegraphics[width=6cm]{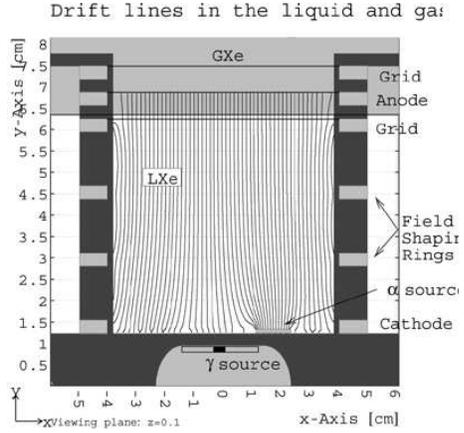} 
   \caption{Electric field distribution for the dual phase prototype.}   
   \label{maxwell} 
  \end{center} 
\end{figure}

To study the electric field, calculations were carried with the Maxwell and Garfield computer programs. The result is shown in Fig. \ref{maxwell}. 
At the drift field of 1 kV/cm, the loss of charges due to drift lines ending-up on the PTFE walls and bottom is minimal. The calculation also verified that the presence of the $^{210}$Po source disk, soldered to the cathode grid wires, does not significantly distort the field lines to affect the drift of ionization electrons produced by the external gamma-ray source.

Seven Hamamatsu R9288 PMTs are installed on the top flange, resting on a PTFE
support. We used LEDs to measure the PMTs gain and single photoelectron response. Additional PMTs or a CsI photocathode will be used in subsequent tests
to increase light collection. The chamber was successfully operated during
several long runs, lasting up to 2 weeks, to test the cryogenics system
reliability, the efficiency of the gas purification/recirculation and to
optimize the dual phase response with both gamma-rays and alpha particles. 

The primary scintillation light from LXe was recorded by a charge integrating 
ADC with a 12 bit resolution. A CAMAC waveform digitizer with 8 bit resolution
was used to record the secondary scintillation light produced in
the amplification gap by the ionization electrons extracted from the liquid to the gas. The coincidence of more than one PMT signals
was required to create a trigger. We are upgrading the DAQ system to digitize
both primary and secondary light pulses.

\subsection{Purification and Recirculation System}
\begin{figure}[htbp] 
  \begin{center} 
   \includegraphics[height=6cm, width=5cm]{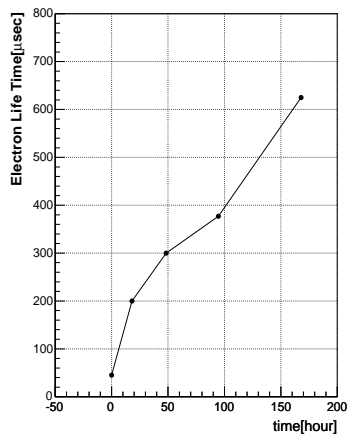} 
   \caption{The electron life time as a function of purification time. Time zero indicates the start of the recirculation.}    
   \label{fig:lifetime} 
  \end{center} 
\end{figure}

To meet the stringent LXe purity requirement for this type of detector, we have developed a gas purification and recirculation system capable to remove electronegative
impurities from the liquid filled detector: xenon gas is continuously extracted
from the detector and circulated through a high temperature getter \citep{SAES}, before being
re-condensed \citep{Mihara:04}. The charge collection efficiency
depends on the electron lifetime in LXe. We have achieved a
lifetime longer than 500 \(\mu s\) after a few days of continuous purification (see Fig.
\ref{fig:lifetime}). Details of the purification/recirculation system are described elsewhere\citep{Aprile:04}.
 
\subsection{Discrimination of Gamma-Rays and Alpha Particles}

\begin{figure}[htbp] 
  \begin{center} 
   \includegraphics[width=5cm]{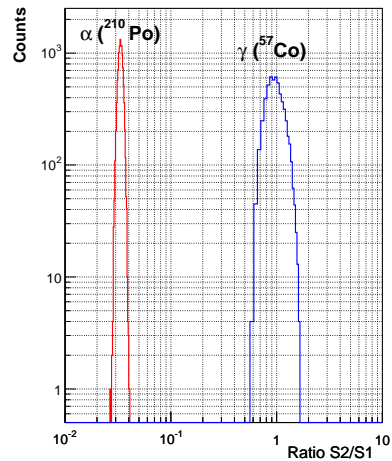} 
   \caption{The ratio $S2/S1$ for 122keV gamma rays and
   5.3 MeV alpha particles. The distributions are normalized to the peak
   from gamma rays.}    
   \label{fig:ratio} 
  \end{center} 
\end{figure}

Fig. \ref{fig:ratio} shows the ratio of light yields $S1$(primary) and
$S2$(secondary) measured with the dual phase prototype, under simultaneous
irradiation with 122 keV gamma rays from $^{57}$Co and with 5.3 MeV alpha particles from
$^{210}$Po. The detector was operated at 1kV/cm in the drift region and
10 kV/cm in the gas region. The ratio is normalized to that of gamma
rays. The alpha and electron recoils are clearly separated by a factor of about
30. This factor can be even larger, if we account for the fraction of primary
light produced by alpha recoils which is absorbed by the $^{210}$Po source disk.

\subsection{Energy Detection Threshold}

The energy detection threshold of the current prototype is limited by the poor
primary light collection efficiency, mainly due to total reflection of
scintillation light at the liquid-gas interface. From Monte Carlo simulations the light collection
efficiency is 0.3 p.e./keV, consistent with experimental
results obtained with this detector. To increase light collection and lower the minimum energy threshold, we need to cover the bottom of the detector with light sensors. In the original
XENON proposal \citep{NSF:01}, we suggested to use a CsI photocathode at the bottom of the LXe drift volume. Simulations show that with the cathode covered by a thin CsI layer, the
primary light collection efficiency of the current prototype would increase to 7.7 p.e./keV, thanks to the high QE and large
coverage of the CsI photocathode \citep{CsI}. The light collection efficiency can also be
improved by placing PMTs below the cathode. With 16 Hamamatsu
R8520 1'' square PMTs on the bottom, we estimate a light collection efficiency of 2.6
or 2.3 p.e./keV, with or without the seven top PMTs. This level is sufficient to
achieve the low energy
detection threshold required for a sensitive WIMP search, and will be our
next upgrade of this prototype prior to realizing the XENON10 module for tests underground.

\section{Background Simulations}
\label{}

The XENON10 dark matter detector prototype, will be deployed at a deep site by the end of 2005.
For a WIMP of 100 GeV, the sensitivity goal for XENON10 is a normalized
cross-section of $2\times10^{-44}$ cm$^2$.
This corresponds to an interaction rate of $\sim$16 events/10 kg/yr,
and a low energy WIMP differential event rate of $\sim7\times10^{-4}$/keVee/kg/day,
for a detection threshold of 8 keVee (16 keV nuclear recoil). A nuclear recoil quenching factor of 50\% is used, to reflect the high field operation ($>$ 1 kV/cm) of the detector.

In the Monte Carlo simulations of the XENON10 background rate, we use this sensitivity as the goal in defining upper limits for acceptable backgrounds, although in many cases the projected background rates are well below this level. The sensitivity goal for the subsequent XENON100 module is estimated to be a factor 10 better than XENON10. In this paper we focus on the XENON10 projections.

The overall background event rate in the detector is contributed by both internal and external sources of gamma rays and neutrons. The total gamma interaction rate in the fiducial volume needs to be $<$0.14 evts/keV/kg/day. This goal is based on a projected discrimination (electron versus nuclear recoils) of at least 99.5\% using the ratio of primary to secondary scintillation light yield.

\begin{figure}[htbp] 
  \begin{center} 
   \includegraphics[width=6.5cm]{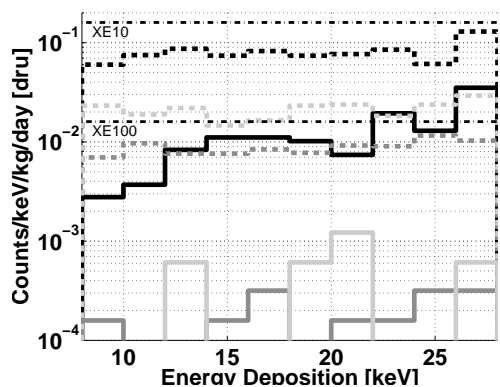} 
   \caption{Monte Carlo event rates in XENON10 arising from 7$\times$inner PMTs (black), 16$\times$outer PMTs (dark gray) and stainless pressure vessel (light gray). The dashed lines are the raw rates, and the solid lines show rates after rejecting multiple site events in the inner LXe, and after the anti-coincidence with the outer LXe shield. The horizontal dot-dash lines indicate the goals for gamma event rates for the XENON10 and XENON100 modules.}
     \label{bghist}
  \end{center} 
\end{figure}

The model of the XENON10 prototype used in the simulations consists of a
cylindrical LXe volume with 17.5 cm diameter and 15 cm depth. The LXe of the
fiducial volume is surrounded by an additional LXe layer, 5 cm
thick, operated as active anticoincidence shield. The number of PMTs
viewing the inner target and the outer shield are 7 and 16, respectively. The U/Th/K activity of the PMTs is assumed to be 13/4/60 mBq/PMT. The numbers reflect the measurements reported to-date by Hamamatsu. We note that the company continues to optimize the choice of materials used in the PMTs construction and expects to further reduce the activity level in the near future.
We also note that the high K activity is not a concern because it contributes
$<$3\% to the low energy gamma background, relative to U/Th of the same
activity. A common vessel made of stainless steel is assumed to contain both the
target and shield LXe. Monte Carlo results show that steel (62 kg) at a level of
activity of 30 mBq/kg (assumed  dominated by $^{60}$Co) will allow us to
comfortably beat ($>100\times$) the target gamma rate in the inner fiducial
region. The result shown in Fig. \ref{bghist} includes event rejection by the active anti-coincidence LXe shield and by requiring only single-sited events in the fiducial volume, as expected from WIMP interactions.

Fig. \ref{bghist}  summarizes the low energy spectra
(evts/keVee/kg/day) of gamma events from the inner and outer PMTs and the stainless steel
pressure vessel. $^{85}$Kr contamination in the Xe will produce a flat 
spectrum in the energy region shown in Fig.  \ref{bghist}
at a level of 1 evts/keVee/kg/day/30 ppb Kr. Kr will be removed from the Xe to levels well below $\ll$1 ppb using charcoal column separation technology being developed by the collaboration. 

The gamma activity from the underground cavern will be attenuated 
using a 22 cm Pb shield. Pb with an activity of $^{210}$Pb of 6 Bq/kg can be 
used for the inner liner ($\sim$5 cm)
of the Pb shield as its effect is reduced by the 5 cm LXe shield.
 
The ($\alpha$,n) neutrons, in the energy range 0.1--8 MeV, arising from U/Th $\alpha$'s in the rock can be attenuated below the
desired level by 60 cm of polyethylene moderator. Although the
underground site has not yet been selected, this projection assumes a
representative flux of 
$4\times10^{-6}$ /cm$^2$/s. 
The contribution to the
neutron background from the internal U/Th content within the shield has also been simulated and shown to be comfortable subdominant. 

Cosmic ray muons contribute electromagnetic backgrounds through
direct interaction in the detector and surrounding shield, and also
through the generation of neutrons in the shield and surrounding
rock. 
A 2" plastic scintillator veto completely surrounding the
Pb/polyethylene shield will be able to tag muons entering the entire
shield volume with $>\sim$99\% efficiency, decreasing the muon associated signals to sub-dominant level at depths $>\sim$2000 mwe. 
High energy neutrons (10-2000 MeV) created in the rock by muons are
not significantly attenuated by the moderator shield, and generate
high multiplicity events within the Pb shield. Depth is the most
effective way to decrease the signal from these "punch-through"
neutrons. Simulations show that the event rate from these punch-through neutrons is reduced below the XENON10 goal at a depth larger than 
~2000 mwe. A deeper location ($>\sim$3700 mwe) will be required to 
achieve the projected sensitivity of XENON100.

\section{Acknowledgment}
This work is supported by a grant from the National Science Foundation to the
Columbia Astrophysics Laboratory (Grant No. PHY-02-01740). The Brown University group is also supported by an Outstanding Junior 
Investigator award from the Department of Energy.
 One of the authors
(P. Majewski) acknowledges the support by the North Atlantic Treaty Organization
under a grant awarded in 2003.





\begin{thebibliography}{}


\bibitem[Aprile(1994a)]{CsI} Aprile, E., et al., Nucl. Inst. Meth. A 338 (1994a) 328.
\bibitem[Aprile(1994b)]{Aprile:94} Aprile, E., et al., Nucl. Inst. Meth. A 343
  (1994b) 129.
\bibitem[Aprile(2004)]{Aprile:04} Aprile, E., et al., IEEE NSS MIC SNPS
			       RTSD (2004) (submitted)
\bibitem[CDMS Collaboration(2004)]{CDMS:04} CDMS Collaboration, astro-ph/0405033.
\bibitem[Chardin(2003)]{Chardin:03} Chardin, G., astro-ph/0306134.
\bibitem[DAMA Collaboration(2003)]{DAMA:03} DAMA Collaboration, Riv. N. Cim. 26
  n.1 (2003) 1-73, also at astro-ph/0307403.
\bibitem[Freedman et al.(2003)]{Freedman:03} Freedman, W.L. and Turner, M.S.,
  Rev. Mod. Phys. 75 (2003) 1433-1447.
Ansoft.
\bibitem[Mihara(2004)]{Mihara:04} Mihara, S., et al., Cryogenics, 
			       44 (2004) 223-228.
\bibitem[Munoz(2003)]{Munoz:03} Munoz, C., hep-ph/0309346.
\bibitem[SAES()]{SAES} SAES Pure GAS, Inc. http://www.saesgetters.com/. 
\bibitem[XENON Collaboration(2001)]{NSF:01}  XENON Collaboration, NSF proposal
  number 0201740, 'XENON: A Liquid Xenon Experiment for Dark Matter', proposal
  submitted to NSF, Particle and Nuclear Astrophysics in Sep. 2001.
\bibitem[XENON Collaboration(2001)]{XENON:01} XENON Collaboration, p.165, Proceedings
  of the International Workshop on Techniques and Applications of Xenon
  Detectors, ICRR, Univ. of Tokyo, Kashiwa, Japan, held December 2001,
  also at astro-ph/0207670.
\bibitem[Yamashita(2003)]{Yamashita:03} Yamashita, M., et al., Astropart. Phys.,
  20 (2003) 79-84.  
\bibitem[Yamashita(2004)]{Yamashita:04} Yamashita, M., et al.,
			       Nucl. Inst. and Meth. A (2004, in press).


\end{thebibliography}
\end{document}